\documentclass{emulateapj}

\usepackage{natbib}
\usepackage{multirow}
\usepackage{hhline}
\slugcomment{Accepted for publication in The Astrophysical Journal.\begin{tiny}\copyright 2009. The American Astronomical Society\end{tiny}}

\shorttitle{Low-mass stars in dense DM halos}
\shortauthors{Casanellas, J. \& Lopes, I.}

\begin{document}


\title{The formation and evolution of young low-mass stars within halos with high concentration of dark matter particles}


\author{Jordi Casanellas\altaffilmark{1,3}, Il\'\i dio Lopes\altaffilmark{1,2,4}}

\altaffiltext{1}{Centro Multidisciplinar de Astrof\'{\i}sica, Instituto Superior T\'ecnico, Av. Rovisco Pais, 1049-001 Lisboa, Portugal} \altaffiltext{2}{Departamento de F\'\i sica,
Universidade de \'Evora, Col\'egio Ant\'onio Luis
Verney, 7002-554 \'Evora - Portugal} 
\altaffiltext{3}{E-mail: jordicasanellas@ist.utl.pt}  \altaffiltext{4}{E-mail: ilidio.lopes@ist.utl.pt}


\begin{abstract} 
The formation and evolution of low-mass stars within dense halos of dark matter (DM) leads to evolution scenarios quite different from the classical stellar evolution. As a result of our detailed numerical work, we describe these new scenarios for a range of DM densities on the host halo, a range of scattering cross sections of the DM particles considered, and for stellar masses from 0.7 to 3 M$_{\odot}$. For the first time, we also computed the evolution of young low-mass stars in their Hayashi track in the pre-main sequence phase and found that, for high DM densities, these stars stop their gravitational collapse before reaching the main sequence, in agreement with similar studies on first stars. Such stars remain indefinitely in an equilibrium state with lower effective temperatures ($\vert\Delta T_{eff}  \vert>10^{3}\;$K for a star of one solar mass), the annihilation of captured DM particles in their core being the only source of energy. In the case of lower DM densities, these proto-stars continue their collapse and progress through the main sequence burning hydrogen at a lower rate. A star of 1$\;$M$_{\odot}$ will spend a time greater than the current age of the universe consuming all the hydrogen in its core if it evolves in a halo with DM density $\rho_{\chi}=10^{9}\;$GeV$\;$cm$^{-3}$. We also show the strong dependence of the effective temperature and luminosity of these stars on the characteristics of the DM particles and how this can be used as an alternative method for DM research.
\end{abstract}





\section{Introduction}
 
Modern observational Cosmology has revealed a more complex and unknown Universe than previously expected. The current cosmological observations of the baryon acoustic oscillations, distance measurements by means of Type Ia supernovae, cosmic microwave background and primordial light elements abundances, are all observations that confirm that the standard cosmological model is a good description of our Universe undergoing an accelerated expansion
\citep{art-Spergeletal2007,art-Komatsuetal2009}. We now know that only about 5 \% of all matter in the Universe is regular visible matter, usually known as baryons. The rest is very likely to be in the form of unknown particles that have not yet been detected, even though we are able to observe its effect on the formation of cosmological structures, usually known to us as dark matter (DM) and dark energy. The Universe is composed of 20\% of DM, particles that do not interact with the electromagnetic field, but whose presence can be inferred from gravitational effects on visible matter. The remaining 75\% is known as dark energy. This component is very homogeneous and not very dense, and is known to have interacted through any of the fundamental forces other than gravity. In this paper we investigate the impact of the DM component on the evolution of stars.
 
There is an abundant number of particle physics models providing candidates for DM particles. Among others, the supersymmetric standard model (SUSY) is one of the best studied candidates for physics beyond the Standard Model \citep{rev-BertoneHS2005}. Dark matter particles must be massive particles, electrically neutral and, very likely, non-coloured. A lightest supersymmetric particle (LSP), such as the neutralino, is the favourite SUSY particle among the particle physics community. The LSP belongs to a generic family of neutral massive particles, usually known as Weakly Interacting Massive Particles (WIMPs), that are by definition the best DM particle candidates. WIMPs interact through gravity and possibly through the weak nuclear force, and very likely through no other interactions stronger than the weak force. 

Despite their weak interactions, WIMPs can lead to significant changes in the formation and evolution of stars, provided they have a sizeable scattering cross section with baryons. Concerning WIMPs' interaction with baryons, the consequence is twofold:  WIMPs can affect a star by annihilating among themselves into standard model particles in its core, providing a source of energy additional to standard nuclear energy \citep{art-SalatiSilk1989}. A second way WIMPS can influence stellar structure is by providing an additional mechanism of heat transport inside the star \citep{art-BouquetSalati1989}. This can reduce the local temperature gradient, potentially inhibiting convection and enhancing the pulsation of horizontal branch stars \citep{art-Dearbornetal1990}. In the case of the Sun, the current values favoured by WIMP self-annihilation cross section, WIMP-nucleus scattering cross sections and local DM density indicate that this effect is not significant for the evolution of our star. 

The study of the effects of DM particles accretion on stars is an alternative approach to investigate the properties of such particles. These effects were first studied in the Sun by \citet{art-SpergelPress1985} and later on by \citet{art-LopesBS2002} and \citet{art-Bottinoetal2002}. Recently, \citet{art-MoskalenkoWai2007} for white dwarfs, \citet{art-Scottetal2007} for low-mass stars, and \citet{let-Spolyaretal2008} for the first generation of stars, showed that, if embedded in halos with high DM densities, stars can be fuelled only by the energy from DM annihilation. Many authors confirmed and developed these results. DM capture and annihilation was studied for white dwarfs and neutron stars by \citet{art-BertoneFairbairn2008} and \citet{art-Kouvaris2008}. The same was done for first stars by \citet{art-Iocco2008}, \citet{art-Taosoetal2008}, \citet{art-FreeseSpolyarAguirre2008}, and \citet{art-YoonIoccoAkiyama2008}, and their implications for cosmic reionization and the pair production supernova rate were studied by \citet{art-SchleicherBK2008} and \citet{art-Iocco2009}, respectively. In the case of first stars formed at the center of DM minihalos, DM annihilation may become the primary source of energy during the adiabatic contraction regime of the DM halo, counteracting the gravitational collapse even before DM capture becomes efficient \citep{let-Spolyaretal2008, art-FreeseGS2008, art-FreeseBSG08, art-Natarajan2009, art-RipamontiIoccoetal2009}. When the star runs out of its original DM, which takes about a million years \citep{art-FreeseReview2009}, this stalling phase ends and the collapse continues until capture replenishes the fuel for DM annihilation \citep{art-Ioccoetal2008}.

Regarding low-mass stars, \citet{art-Scottetal2009} carried out an extensive study on how the stars located at the galactic center may be affected by the dense DM densities which are expected to exist there. They assumed that these stars were already formed in a scenario without DM. Therefore, their evolution on a DM halo was considered from the ZAMS. In addition to that, we also took into account the capture of DM particles from the collapse of the protostar, as \citet{art-Ioccoetal2008} did for first stars. The influence of DM in that early stage may have dramatic consequences in the forming star, as will be shown.

The paper is organised as follows: in section \ref{sec-thecode} there is a short description of the stellar evolution code used here. The process of accretion of annihilating DM particles in stars is briefly reviewed in section \ref{sec-ADM}. The different evolution paths found for low-mass stars evolving within DM halos are discussed in section \ref{sec-Evol}. In section \ref{sec-Stellar} we discuss the implication of our results to observational stellar astrophysics. Finally, in section \ref{sec-conclusion}, we present a brief summary of the results and discuss the implication of such results for DM research.
 
\section{The stellar evolution code}
\label{sec-thecode}

The models computed in this work were made using the stellar evolution code CESAM \citep{art-Morel1997}. The CESAM code is a consistent set of programs and routines which performs calculations of 1D quasi-static stellar evolution including diffusion, rotation and mass loss. CESAM computes structure equilibrium models by collocation method based on piecewise polynomials approximations projected on their B-spline basis, the evolution of the chemical composition is solved by stiffly stable schemes of orders up to four, the solution of the diffusion equation employs the Petrov-Galerkin scheme \citep{art-Morel1997}. The code determines the evolution of a
star by the integration of the set of conservation structure equations, coupled with a set of nuclear reactions describing the nucleosynthesis of chemical elements and the production of energy, using an adaptive space-time mesh, i.e., the code chooses an optimal step in time and space by computing the rate of variation of the equilibrium quantities. CESAM allows calculations of stellar models with various physical assumptions, physical data, external boundary conditions, numerical methods and numerical accuracy; in this work the accuracy is set to $10^{-5}$. This code has the ability to compute the evolution of stars from the pre-main sequence up until the beginning of the $^4$He burning cycle for various stellar masses, and for a range of metallicities (0.0004$<\;$Z$\;<$0.04, in the present configuration). The initial chemical abundances are the solar ones \citep{art-AsplundGrevesseSauval2005} and the initial metallicity, unless stated otherwise, is assumed to be Z=0.014, similar to that of the Sun.

The microscopic physics, including equation of state, opacities, nucleosynthesis, microscopic diffusion and chemical abundances, is very refined in this stellar evolution code, since this part of the code was tested in the case of the Sun against helioseismic data. The CESAM evolution code is well established in the solar and stellar physics comunity, being used either to compute solar models \citep{art-CouvidatTuChi2003, art-Berthomieu1993} or models of other stars (among others: PMS$\delta$Scuti star V346 Ori \citep{art-Bernabeietal2009}, $\beta$Chepei star $\nu$ Eridani \citep {art-Suarezetal2009}, 0.8-8M$_{\odot}$ PMS and MS stars \citep{art-Marques2008}, Vega-like stars $\alpha$PsA, $\beta$Leo, $\beta$Pic, $\varepsilon$Eri and $\tau$Cet \citep{art-DiFolco2004}, $\alpha$CMi star Procyon A \citep{art-Kervella2004}, solar-like 0.8-1.4M$_{\odot}$ stars and PMS stars \citep{art-PiauTurkChieze2002}, $\alpha$Cen A\&B stars \citep{art-Thevenin2002} and MS stars with convective overshooting \citep{art-Audardetal1995}).

We modified this code to take into account the impact of DM particles in the evolution of low-mass stars. In particular, we have already used this code to study the evolution of the Sun within a halo of dark matter \citep{art-LopesBS2002,art-LopesSH2002,let-LopesSilk2002}; we predicted the results of two groups of observables: solar neutrinos and helioseismology data (including the sound speed profile).

\section{Accumulation of Dark Matter Particles Inside Stars}
\label{sec-ADM}

WIMPs travel through stars, where they experience scattering off the nuclei that they encounter in the stellar cores. Although most WIMPs travel right through the star without suffering any type of interaction, some of them will scatter off nuclei losing energy. If they lose enough energy, they won't be able to escape the gravitational field of the star anymore. The number of WIMPs trapped inside the star is measured by the capture rate $C_\chi$. 
The capture rate is proportional to the WIMP scattering cross section off nuclei $\sigma_\chi$ and the dark matter density of the halo $\rho_\chi$. It is inversely proportional to the WIMP mass $m_\chi$ and to the WIMP dispersion velocity $\bar{v}_\chi$.  Capture rates of WIMPs were first calculated by \citet{art-PressSpergel1985} in the case of the Sun, by \citet{art-Gould1987} for generic massive bodies and for the earth in particular, and by \citet{art-BouquetSalati1989} for main sequence stars. We computed the capture rate using equation 2.31 of \citet{art-Gould1987}:
\begin{eqnarray}
C_\chi=\sum_{i}\left(\frac{8}{3\pi}\right)^{1/2}\hspace{-3mm}\sigma_{i}\frac{\rho_{\chi}}{m_{\chi}}\bar{v}_{\chi}\;\frac{x_{i}M_{\star}}{A_{i}m_p}\;\frac{3 v_{esc}^2}{2 \bar{v}_{\chi}^2}\;\zeta,
\label{eq_capture}
\end{eqnarray}
where $v_{esc}=\sqrt{2GM_{\star}/R_{\star}}$ is the escape velocity, $M_{\star}$ and $R_{\star}$ are the mass and radius of the star, $m_p$ is the proton mass, and $x_i$ and $A_{i}$ are, respectively, the mass fraction and the number of nucleons of element $i$. We included the contributions from 13 nuclei: H, $^4$He, $^{12}$C, $^{14}$N, $^{16}$O, $^{2}$H, $^{3}$He, $^{7}$Li, $^{7}$Be, $^{13}$C, $^{15}$N, $^{17}$O and $^{9}$Be, the abundances of which are followed by our code. For all of them, we computed their spin-independent (SI) interaction with WIMPs (coherent scattering), which scales as the fourth power of the nucleus mass number, leading to a scattering cross section $\sigma_{i}=A_i^4\sigma_{\chi,SI}$. For hydrogen, we also took into account the contribution of the spin-dependent (SD) interaction (incoherent scattering), which lead to $\sigma_{H}=A_H^4\sigma_{\chi,SI}+\sigma_{\chi,SD}$. The last term in equation \ref{eq_capture}, $\zeta$, is the product of the suppression factors due to the motion of the star through the halo $\xi_\eta(\infty)$, the mismatch between WIMP and nuclei masses, and the dimensionless gravitational potential averaged over the star $<\widehat{\phi}>$, all present in the original expression of \citet{art-Gould1987}. We evaluated these factors for a WIMP mass m$_{\chi}=100\;$GeV, a velocity of the star v$_{\star}=220\;$km$\;$s$^{-1}$ and a WIMP dispersion velocity $\bar{v}_\chi=270\;$km$\;$s$^{-1}$, and found that, for these values and for all stars studied here, we can approximate $\zeta$ to the order of unity, similarly to other authors \citep{art-MoskalenkoWai2007, art-BertoneFairbairn2008, art-FreeseSpolyarAguirre2008}. This approximation retains the main aspects of WIMPs' capture while it simplifies the calculations and reduces the time of computation. Furthermore, the difference in values between this capture rate and others (see table \ref{tab_comp}) is relatively small. We confirmed that the overall conclusions of this paper will not be affected by the accuracy of this capture expression.

\begin{table}[htbp]
\centering
\begin{tabular}{cccl}
\hline
\hline
Mass & $\log\rho_{\chi}$ & C$_{\chi}$ & \multirow{2}{*}{Reference}\\
($M_{\odot}$) & (GeV cm$^{-3}$) & (s$^{-1}$) & \\
\hline
&&&\\
 0.8 & 9 & $0.6\cdot10^{34}$ & \citet{art-Scottetal2009}$^{\dag}$ \\
 & & $0.9\cdot10^{34}$ & \citet{art-Fairbairnetal2008} \\
 & & $1.7\cdot10^{34}$ & \citet{art-FreeseSpolyarAguirre2008}$^{\ddagger}$ \\
 & & $1.6\cdot10^{34}$ & Casanellas \& Lopes$^{\dag}$ \\
&&&\\
 1 & 10 & $0.8\cdot10^{35}$ & \citet{art-Scottetal2009}$^{\dag}$ \\
 & & $0.9\cdot10^{35}$ & \citet{art-Fairbairnetal2008} \\
 & & $1.1\cdot10^{35}$ & \citet{rev-BertoneHS2005} $^{\ast}$ \\
 & & $1.7\cdot10^{35}$ & \citet{art-FreeseSpolyarAguirre2008}$^{\ddagger}$ \\
 & & $2.9\cdot10^{35}$ & \citet{art-MoskalenkoWai2006}$^{\ddagger}$ \\
 & & $1.6\cdot10^{35}$ & Casanellas \& Lopes$^{\dag}$ \\
&&&\\
 & 9 & $1.0\cdot10^{34}$ & \citet{art-Scottetal2009}$^{\dag}$ \\
 & & $1.1\cdot10^{34}$ & \citet{art-Fairbairnetal2008} \\
 & & $1.1\cdot10^{34}$ & \citet{rev-BertoneHS2005}$^{\ast}$ \\
 & & $2.0\cdot10^{34}$ & \citet{art-FreeseSpolyarAguirre2008}$^{\ddagger}$ \\
 & & $1.9\cdot10^{34}$ & Casanellas \& Lopes$^{\dag}$ \\
&&&\\
 & 8 & $0.8\cdot10^{33}$ & \citet{art-Scottetal2009}$^{\dag}$ \\
 & & $1.1\cdot10^{33}$ & \citet{art-Fairbairnetal2008} \\
 & & $1.1\cdot10^{33}$ & \citet{rev-BertoneHS2005}$^{\ast}$ \\
 & & $2.1\cdot10^{33}$ & \citet{art-FreeseSpolyarAguirre2008}$^{\ddagger}$ \\
 & & $1.9\cdot10^{33}$ & Casanellas \& Lopes$^{\dag}$ \\
&&&\\
 2 & 10 & $2.3\cdot10^{35}$ & \citet{art-Fairbairnetal2008} \\
 & & $4.4\cdot10^{35}$ & \citet{art-FreeseSpolyarAguirre2008}$^{\ddagger}$ \\
 & & $4.5\cdot10^{35}$ & \citet{art-Scottetal2009}$^{\dag}$ \\
 & & $4.1\cdot10^{35}$ & Casanellas \& Lopes$^{\dag}$ \\
&&&\\
 3 & 10 & $4.5\cdot10^{35}$ & \citet{art-Fairbairnetal2008} \\
 & & $8.5\cdot10^{35}$ & \citet{art-FreeseSpolyarAguirre2008}$^{\ddagger}$ \\
 & & $8.1\cdot10^{35}$ & Casanellas \& Lopes$^{\dag}$\\
\hline
\end{tabular}
\caption{Comparison of different capture rates in the literature, considering different stellar masses and DM halo densities $\rho_{\chi}$. In all cases: m$_{\chi}=100$ GeV, $\sigma_{\chi,SD} = 10^{-38} $cm$^{2}$. The metallicity of the stars goes as follows: $^{\dag}$:Z=0.01, $^{\ddagger}$:Z=0 and $^{\ast}$:Z=0.018. }
\label{tab_comp}
\end{table}

The values of the scattering cross sections used in our simulations, $\sigma_{\chi,SD}$ from $10^{-40}\;$cm$^{2}$ to $10^{-37}\;$cm$^{2}$ and $\sigma_{\chi,SI} = 10^{-44}\;$cm$^{2}$, are consistent with the experimental bounds given by direct detection experiments (CDMS \citep{art-CDMS_SD2006}, XENON10 \citep{art-XENON10_SD2008}, NAIAD \citep{art-NAIAD2005}, PICASSO \citep{art-PICASSO2005}, COUPP \citep{art-COUPP2008} for $\sigma_{\chi,SD}$, and CDMS-II \citep{art-CDMSII_SI2009}, XENON10 \citep{art-XENON10_SI2008}, CRESST-II \citep{art-CRESST-II2005} and EDELWEISS \citep{art-EDELWEISS-I2005} for $\sigma_{\chi,SI}$). Only one indirect detection experiment (Super-Kamiokande \citep{art-SuperKAMIOKANDE2004}) predicted an upper limit for $\sigma_{\chi,SD}$ below 10$^{-37}$ cm$^2$. The bounds on the WIMP-nucleon scattering cross sections are much less constraining for SD interactions than for SI, due to the presence of the A$^4$ factor in the latter type of interactions. For this reason, the WIMP capture rate was always dominated by SD scattering in all our computations. 

After being captured, WIMPs will then sink to the core of the star, where they can annihilate with one another at an annihilation rate $\Gamma_\chi$.  The total number of WIMPs $N_\chi$ inside the star is then determined by the balance between the capture rate and the annihilation rate. Therefore, the evolution of the number of WIMPs inside the star over time is 
\begin{eqnarray}
\frac{dN_\chi}{dt}=C_\chi-2\Gamma_\chi. 
\end{eqnarray}
The capture and annihilation of WIMPs in the core of a star is a very efficient process; capture and annihilation processes balance each other out in around 100 years \citep{art-SalatiSilk1989} and the system rapidly comes to equilibrium, $\dot{N}_\chi=0$ or $C_\chi=2\Gamma_\chi$. The time scale for the steady state to be reached will be much inferior than the typical evolutionary time scale of a main sequence star. WIMPs will get distributed in the core of the star very rapidly following an approximately thermal internal distribution with a characteristic radius  $r_\chi=\sqrt{3\kappa T_c/2\pi G\rho_c m_\chi}$, where $T_c$ and $\rho_c$ are the central temperature and density of regular baryonic matter inside the star, and $G$ and $\kappa$ are the Newton and Boltzman constants. The density number distribution $n_\chi$ is given by
$n_\chi(r)=N_\chi\pi^{-3/2}r_\chi^{-3}\;e^{-r^2/r_\chi^2}$ 
\citep{art-Dearbornetal1990,art-DearbornGriestRaffelt1991,art-Kaplanetal1991}.

Dark matter particles can transport energy by scattering off nuclei inside the star, thus constituting an alternative mechanism of energy transport \citep{art-BouquetSalati1989b}. This energy transport mechanism by WIMPs is implemented in our stellar evolution code (for a detailed descripton, see \citet{art-LopesSH2002}), although the contribution of such process to the evolution of the star is negligible compared with other effects. More significantly, DM particles in the stellar core provide an extra source of energy \citep{art-SalatiSilk1989}. The energy generation rate $\displaystyle{\varepsilon}_\chi$ due to pair annihilation of DM particles is given by
\begin{equation}
\displaystyle{\varepsilon}_{\chi}(r) = f_\chi  \;m_{\chi}\;n_{\chi}^2(r)\,\rho(r)^{-1}<\sigma_{a}v>,
\label{eq_energy}
\end{equation}
in units of energy per mass per time (in cgs: erg$\;$g$^{-1}\;$s$^{-1}$). It follows that every pair of DM particles captured into the star is instantly converted into additional luminosity. We assume that all products of DM annihilation, except neutrinos, interact either by electromagnetic or nuclear strong forces with the core nuclei, so they have short mean free paths inside the star and these particles rapidly reach the thermal equilibrium. We chose the coefficient $f_\chi=\frac{2}{3}\times2$ to take into account that $\frac{1}{3}$ of the energy is lost in the form of neutrinos that escape from the star, and that each annihilation involves 2 DM particles, due to the assumption that the neutralino is a Majorana particle, i.e. it is its own anti-particle. The coefficient $f_\chi$ has different values in the literature. Its former factor, which quantifies the energy that remains inside the star, could be underestimated in our work: following the recent simulations of \citet{art-Scottetal2009}, the energy loss could be as low as 10\% of the total energy from DM annihilations. Our choice, more conservative, is in agreement with other authors \citep{art-FreeseBSG08, art-Ioccoetal2008, art-YoonIoccoAkiyama2008}. In our simulations we assumed the annihilation cross section to be $<\sigma_a v >=3\cdot10^{-26}\;$cm$^3\;$s$^{-1}$, a value that is fixed by the relic density through the following approximation: $\Omega_{\chi}h^2\approx3\cdot10^{-27}\;$cm$^3\;$s$^{-1}/<\sigma_a v >$ (e.g., \citet{art-ScherrerTurner1986, art-Srednickietal1988, rev-BertoneHS2005}).

\section{Evolution of low-mass Stars within a Dark Matter Halo}
\label{sec-Evol}

\begin{figure}[!t]
\centering
\includegraphics[scale=0.8]{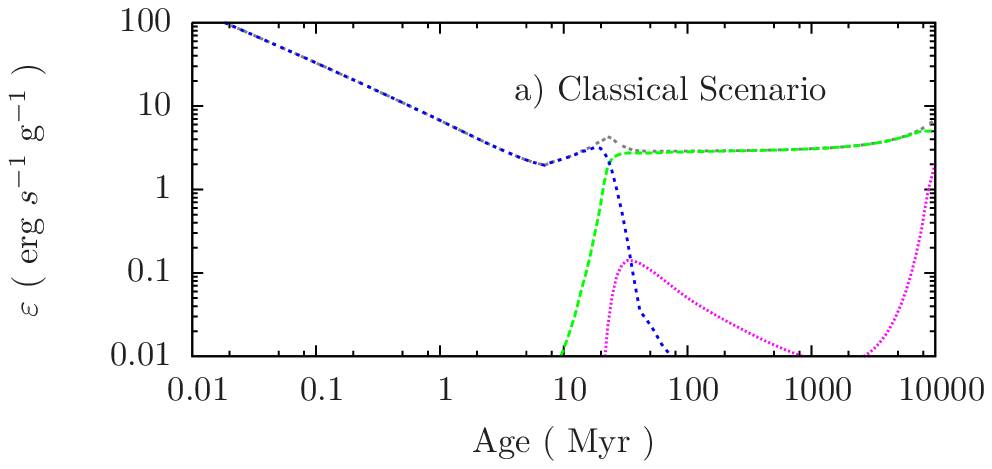}
\includegraphics[scale=0.8]{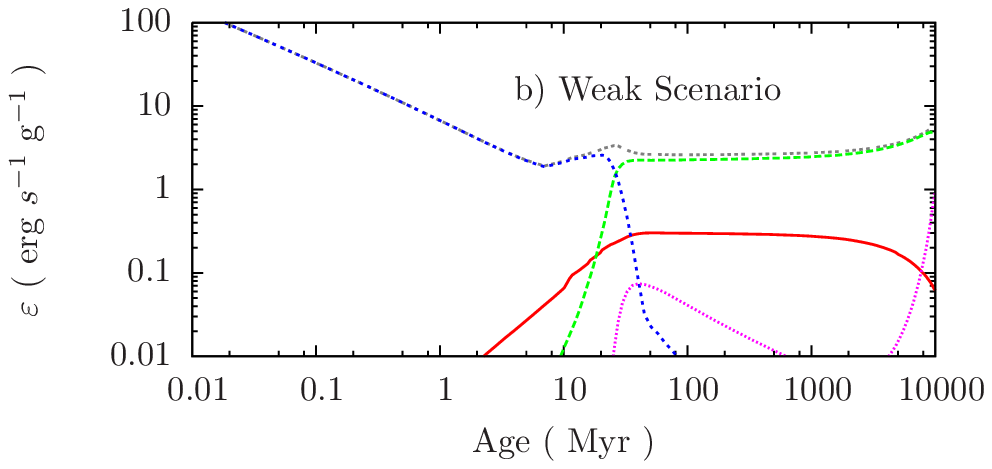}
\includegraphics[scale=0.8]{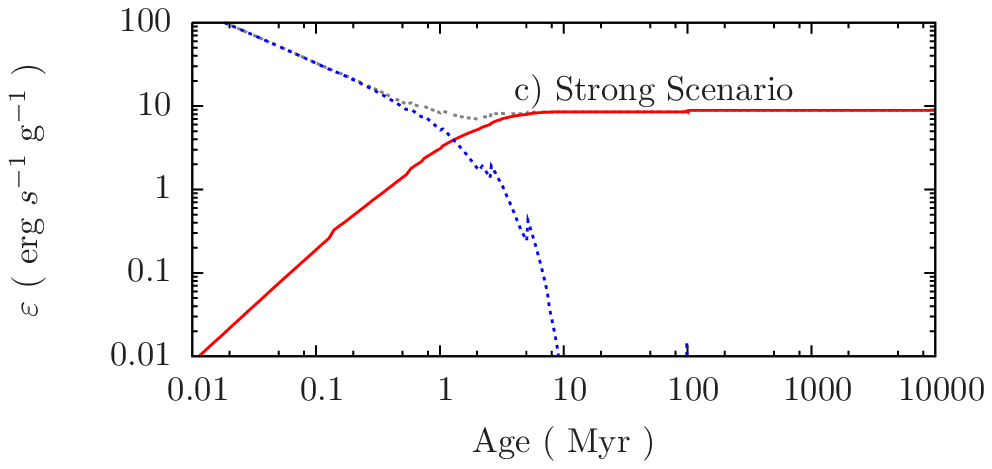}
\caption{Energy rates during the evolution of a $1\;M_{\odot}$ star on the Classical, Weak and Strong cases. The densities of dark matter in the halo are, respectively: 0, $3\cdot10^{8}$ and $3\cdot10^{11}\;$GeV$\;$cm$^{-3}$. Curves are as follows: $\varepsilon_{T}$ (light grey dashed line), $\varepsilon_{\chi}$ (red continuous line), $\varepsilon_{grav}$ (blue dashed line), $\varepsilon_{pp}$ (green long dashed line), $\varepsilon_{CNO}$ (pink dotted line). The DM halo is assumed to be composed by WIMPs with mass $m_\chi=100\;$GeV, spin-dependent scattering cross section, $\sigma_{\chi,SD}=10^{-38}$ cm$^{2}$, and annihilation cross section $<\sigma_a v >=3\cdot10^{-26}\;$cm$^3\;$s$^{-1}$.}
\label{f_3reg}
\end{figure}

We have implemented the effects of the annihilation of DM particles in our stellar evolution code and followed the evolution of low-mass stars since their proto-star phase and throughout the main sequence phase. These stars may experience dramatic changes on their evolution depending upon the amount of DM the star accumulates in its interior. The accretion of DM depends mainly on the particle halo density $\rho_{\chi}$, and also on the WIMP-nucleus spin-dependent scattering cross section $\sigma_{\chi,SD}$. The more accretion of DM particles happens inside the core of the star, the more energy is produced by WIMP pair annihilation. The existence of this new source of energy leads to significantly different scenarios of stellar evolution. Figure~\ref{f_3reg} shows the contribution of the different energy sources to the total energy generation rate, $\displaystyle{\varepsilon_T}$, of a star of 1 M$_{\odot}$. The  evolution of the star depends on the balance between DM energy rate, $\displaystyle{\varepsilon}_\chi$, the thermonuclear energy rate produced by the $pp$ chain, $\varepsilon_{pp}$,  the thermonuclear energy rate produced by the $CNO$ cycle, $\varepsilon_{CNO}$, and the gravitational energy rate, $\varepsilon_{grav}$, produced by the gravitational contraction of the star. Depending upon the amount of DM present in the host halo, we found that stars can experiment quite different evolution paths, which we classified in three distinct cases: Weak, Intermediate and Strong case scenarios.

\subsection{Scenarios of stellar evolution within DM}

\subsubsection{Weak case scenario: Slowly evolving stars}

\begin{figure}[!tb]
\centering
\vspace{0.1cm}
\includegraphics[scale=0.8]{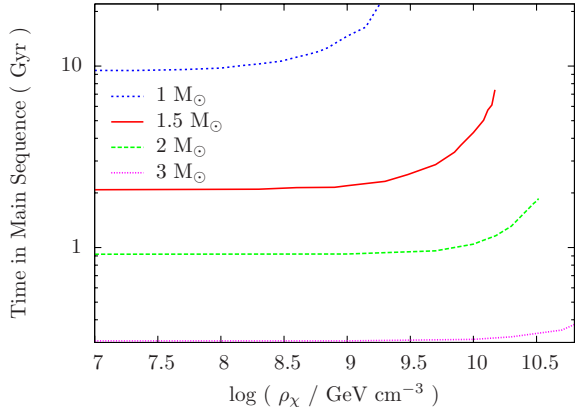}
\caption{Time spent in main sequence (from $\varepsilon_{grav}<1\%\,\varepsilon_{T}$ to $X_{c} < 0.001$) by stars of masses 1, 1.5, 2 and 3$\;$M$_{\odot}$ embedded in dark matter halos of different densities. The DM particles are as described in Figure \ref{f_3reg}.}
\label{f_timeMS}
\end{figure}

Normal stars are self-gravitating systems that most of the time are experimenting a gravitational contraction, leading to an increase in the temperature inside their cores. The gravitational collapse is stopped by an additional source of energy, such as thermonuclear energy produced by the $pp$ chain or $CNO$ cycle in stars on the main sequence phase.
Nevertheless, stars evolving in DM halos can experiment a quite different scenario of evolution. For stars evolving within halos with low DM density $\rho_\chi$, the energy from WIMPs' annihilation is a complementary source to the thermonuclear energy (see Figure~\ref{f_3reg}.b), that slightly delays the gravitational collapse, slowing down the arrival of the Hydrogen burning phase. The equilibrium is reached at a lower central temperature than that of the classical evolution case, leading to a smaller rate of energy produced by thermonuclear reactions $\varepsilon_{pp} + \varepsilon_{CNO}$ (stars will evolve in the weak scenario if their thermonuclear energy accounts for more than 10\% of the total energy in the beginning of the MS). Therefore, the time that a star spends in the main sequence phase is enlarged respect to the classical evolution scenario (see Figure~\ref{f_timeMS}). The more massive the star, the more DM will be necessary to produce the same effects. A star of 1M$_{\odot}$ will stay in the MS for a time greater than the current age of the universe if it evolves in a DM halo of $\rho_{\chi}= 10^{9}\;$GeV$\;$cm$^{-3}$ and $\sigma_{\chi,SD}=10^{-38}\;$cm$^{2}$, while a star of 2$\;$M$_{\odot}$ evolving in the same halo will not be affected. In the case of one solar mass stars, \citet{art-Scottetal2009} obtained the same extension in the main-sequence lifetime for almost identical DM densities on the host halo. On the other hand, for greater masses our results are more conservative, due to the lower WIMP capture rates obtained for $M_{\star}>1\;$M$_{\odot}$. This evolution scenario is qualitatively similar to that predicted for Pop III stars by \citet{art-Taosoetal2008}.

To grasp the role of the metallicity, we computed models with metallicities from Z=0.0004 to Z=0.04 and found that the main differences in the stellar evolution are those already expected in the classical picture; stars with higher metallicities have lower thermonuclear energy production rates and, therefore, extended main sequence lifetimes \citep{art-Schalleretal1992, art-Schaereretal1993, art-LejeuneShaerer2001}. Table \ref{tab_metall} shows the energy rates $\varepsilon_{pp}$ and $\varepsilon_{\chi}$ for stars with metallicities Z=0.0004, 0.001, 0.02, and 0.04 that evolve in a halo with a DM density $\rho_{\chi}=10^9\;$GeV$\;$cm$^{-3}$. Even though $\varepsilon_{\chi}$ is lower in high metallicity stars (due to a lower capture because of the smaller hydrogen mass fraction), its percentage over the total energy is higher given the strong reduction in $\varepsilon_{pp}$. Consequently,
stars with higher initial metallicities will be affected by the energy from DM annihilation at slightly lower DM densities.

\begin{table}[!htbp]
\centering
\begin{tabular}{l l l c c c}
\hline
\hline
\multirow{2}{*}{Z} & \multirow{2}{*}{X$_{in}$} & \multirow{2}{*}{Y$_{in}$} & $\varepsilon_{\chi}$ & $\varepsilon_{pp}$ \\
& & & (erg$\;$g$^{-1}\;$s$^{-1}$) & (erg$\;$g$^{-1}\;$s$^{-1}$) \\
\hline
0.0004	& 0.7584 & 0.2412 & 0.8 (12\%) & 5.8 (88\%) \\
0.001 & 0.756 & 0.243 & 0.7 (12\%) & 5.4 (88\%) \\
0.02 & 0.680 & 0.300 & 0.6 (23\%) & 2.1 (77\%)  \\
0.04 & 0.620 & 0.340 & 0.6 (25\%) & 1.7 (74\%) \\
\end{tabular}
\caption{Energy rates (and its percentage over the total energy) for stars of 1 M$_{\odot}$ with different initial metallicities, evolving in a halo with a DM density $\rho_{\chi}=10^9\;$GeV$\;$cm$^{-3}$, at an age such that their central hydrogen mass fraction is X$_c$=0.60.}
\label{tab_metall}
\end{table}

\begin{figure}[!t]
\centering
	\includegraphics[scale=0.8]{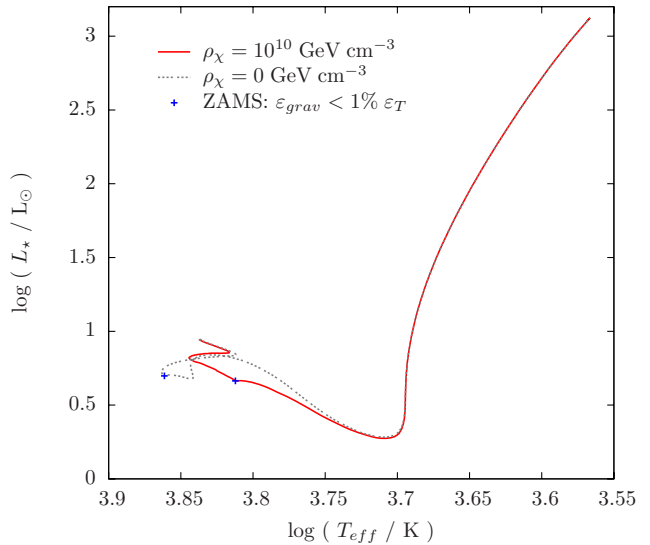}
\caption{HR diagram of the formation and evolution through the main sequence of a star of 1.5$\;$M$_{\odot}$ embedded in halos without DM (grey line) and with a DM halo density $\rho_{\chi} = 10^{10}\;$GeV$\;$cm$^{-3}$ (red line). The DM particles are as described in Figure \ref{f_3reg}.}
\label{f_diffHR}
\end{figure}

The competition between the nuclear burning and the energy from WIMP annihilation leads to another important change respect to the classical scenario. The path that these stars follow on the HR diagram may be significantly altered if there is enough DM in their interior. This can be seen in Figure \ref{f_diffHR}, where we plotted the tracks of a star of 1.5 M$_{\odot}$ evolving in halos without DM and with a DM density $\rho_{\chi} = 10^{10}\;$GeV$\;$cm$^{-3}$. In the latter case, the rates of energy production were, in the beggining of the MS: $\displaystyle{\varepsilon}_\chi \simeq 42\%$, $\varepsilon_{pp} \simeq 55\%$ and $\varepsilon_{CNO} \simeq 3 \%$. In this case, the contribution of $\varepsilon_{\chi}$ rapidly starts to compensate the collapse of the proto-star, stopping it when the star has a larger radius than that of the classical scenario. Consequently, the ZAMS position of these stars shifts to lower effective temperatures. 

The internal structure of the star also experiences some important changes. In the classical scenario, a Sun-like star of one solar mass will develop a small convective core within the radiative interior, which will disappear when the star reaches the main sequence phase. Then, this star will develop a radiative core and a convective envelope on the outer layers. Alternatively, in this new scenario, the central convective zone remains for a longer period during the evolution of the star, because the extra luminosity amount produced by the WIMPs annihilation requires a more efficient mechanism to evacuate this energy from the stellar core. Typically, in a halo of DM particles with $\rho_{\chi}=10^{9}\;$GeV$\;$cm$^{-3}$ and $\sigma_{\chi,SD}=10^{-38}\;$cm$^{2}$, a star of 1$\;$M$_{\odot}$ develops a convective core with a radius of approximately $0.05\;$R$_{\star}$. This radius decreases with time, until it disappears completely at an age of 6 Gyr. Similarly to the Sun, this star conserves its convective external region throughout its evolution. In the case of the present Sun, the external convective region is located  above $0.7\;$R$_{\star}$. The thickness of this external convective layer grows with $\rho_{\chi}$, and it is also conserved in more dense DM halos, as in the case discussed in the next section, the intermediate case scenario.


\subsubsection{Intermediate case scenario: Convective-radiative ``frozen'' stars}

As the ambient density of DM particles $\rho_\chi$ increases, the capture rate inside the star increases too. As a consequence, the energy source resulting from the annihilating DM particles eventually starts to compensate the gravitational energy source, thus keeping the temperature of different core regions below the threshold needed to start the thermonuclear reactions. In this scenario, WIMP pair annihilation becomes the only source of the star's luminosity and $\varepsilon_{\chi}$ is high enough to stop the gravitational collapse at an early stage. Stars in this scenario can live forever without the production of thermonuclear energy.

Since WIMP annihilation occurs in a more centralized region than the nuclear one, at least for very massive DM particles, the temperature gradient is much steeper in the core of the star than it would be otherwise, and the star has the conditions to maintain a convective core for the rest of its life. Outside the convective core, less energy is generated per unit volume than if the nuclear burning was proceeding normally, so the temperature gradient is smaller. The actual temperature is lower than in a normal star, however, it remains high enough to prevent any major increase in opacity, ensuring that energy transport in the region above the core remains radiative. The energy from the core is easily transmitted through this radiative region to the surface of the star. The external envelope of this star is very much similar to a typical young Sun. In the outer layers, the star develops a convective region; due to the rapid temperature drop, some chemical elements such as oxygen, carbon and nitrogen,
fully ionised in the interior, are partially ionised in the most external layers. This increases significantly the radiative opacity, and makes convection the only efficient mechanism of energy transport towards the surface. 


\subsubsection{Strong case scenario: Fully convective ``frozen'' stars}

\begin{figure}[!t]
\centering
\vspace{0.1cm}
\includegraphics[scale=0.8]{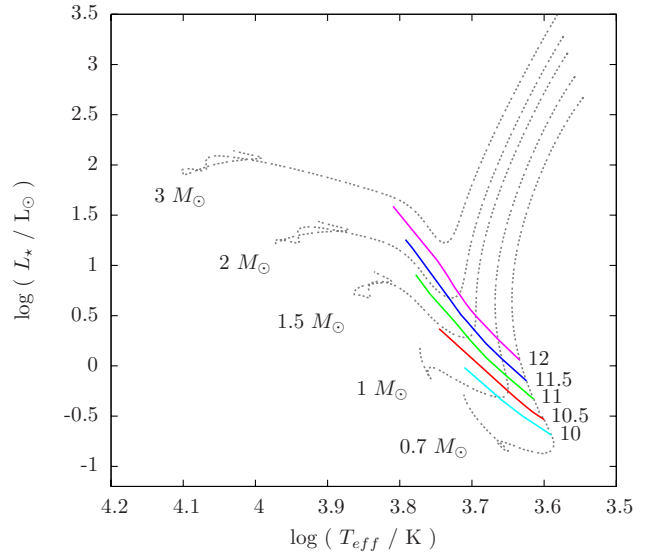}
\caption{Stationary states reached by stars with masses from 0.7 to 3$\;$M$_{\odot}$ when the energy from DM annihilation compensates the gravitational energy during the collapse. These equilibrium positions, where stars will remain for an indefinite time, are plotted for different dark matter halo densities, indicated in units of $\log(\rho_{\chi}/$GeV$\;$cm$^{-3}$) at the side of each line. The grey lines are the classical evolutionary paths, which these stars follow before stopping. The DM particles are as described in Figure \ref{f_3reg}.}
\label{f_statHR}
\end{figure}

At high enough WIMP capture rates, the energy produced by the annihilation of DM is sufficient to fully compensate the gravitational energy (see Figure~\ref{f_3reg}.c). The star's gravitational collapse stops before reaching enough central temperature to begin nuclear fusion, as in the intermediate case. This equilibrium is reached quite early in the formation of these stars, depending upon the value of $\rho_\chi$, as illustrated in Figure~\ref{f_statHR}.  
\begin{figure}[!ht]
\centering
\vspace{0.4cm}
\includegraphics[scale=0.8]{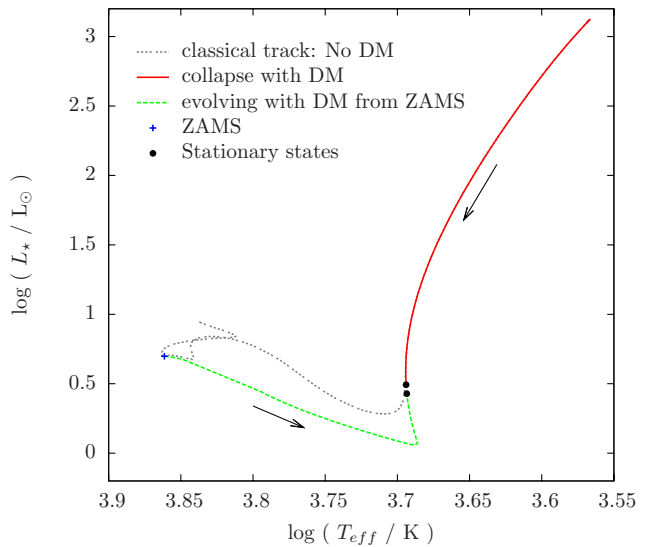}
\caption{HR diagram of a star of 1.5$\;$M$_{\odot}$ evolving in a halo of DM with $\rho_{\chi} = 10^{12}\;$GeV$\;$cm$^{-3}$. The stationary states are new equilibrium states where the star is powered only by energy from DM annihilation. These states are approximately equivalent either if they are reached from the collapse of the protostar (red line), or considering that the star evolves in a halo of DM from the ZAMS (green line). The grey line shows the normal HR track of a 1.5$\;$M$_{\odot}$ star without DM. The DM particles are as described in Figure \ref{f_3reg}.}
\label{f_twosides}
\end{figure} This Figure shows solid grey lines corresponding to the classical evolution tracks of stars from 0.7 to 3$\;$M$_{\odot}$, along with the position on the HR diagram of the early equilibrium states reached by these stars, considering formation scenarios with different WIMP densities $\rho_\chi$. 
These cases are strongly reminiscent of the Hayashi track which young stars follow when travelling along the proto-star phase onto the final stages of their formation towards the main sequence phase. If no extra source of DM energy existed, these stars would shrink in size within the Kelvin-Helmholtz timescale as they radiate away gravitational energy. Alternatively, a constant energy generation in the core by WIMP pair annihilation creates stars that can in principle remain in the same position in the HR diagram for an arbitrarely long time. A star of one solar mass evolving within a halo of DM particles with a scattering cross section $\sigma_{\chi,SD}=10^{-38}\;$cm$^{2}$ and $\rho_{\chi}=3\cdot10^{11}\;$GeV$\;$cm$^{-3}$ completely stops its collapse at the age of $50\;$Myr and remains forever with a stellar radius $R_{\star} = 1.75\;$R$_{\odot}$, effective temperature $T_{eff} = 4555\;$K and luminosity  $L_\star = 1.2\;$L$_{\odot}$ (Cf. Figure~\ref{f_statHR}).

The addition of more WIMPs dramatically increases the central luminosity of these stars, requiring the convective core to grow in order to transport the additional energy to the surface layers. In this scenario, the surface convection zone merges with the inner core and the star becomes fully convective.

\begin{figure}[!t]
\centering
\vspace{0.2cm}
	\includegraphics[scale=0.8]{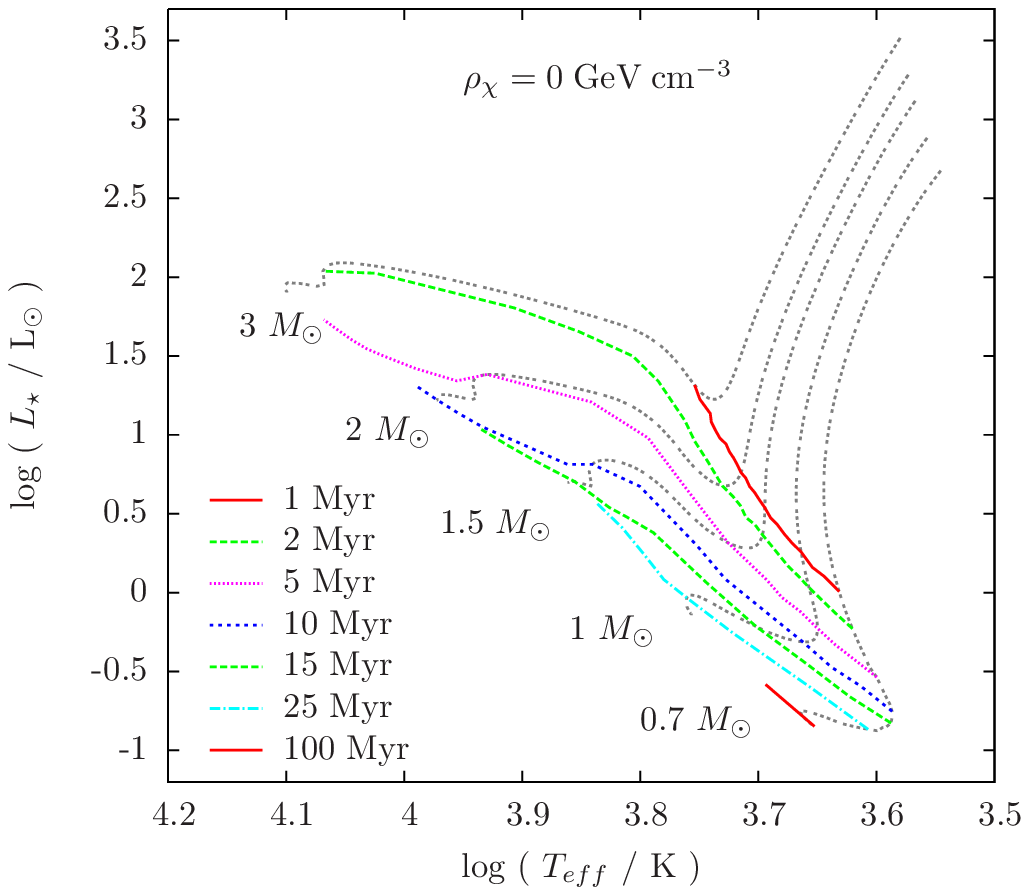}
\vspace{0.7cm} \\
	\includegraphics[scale=0.8]{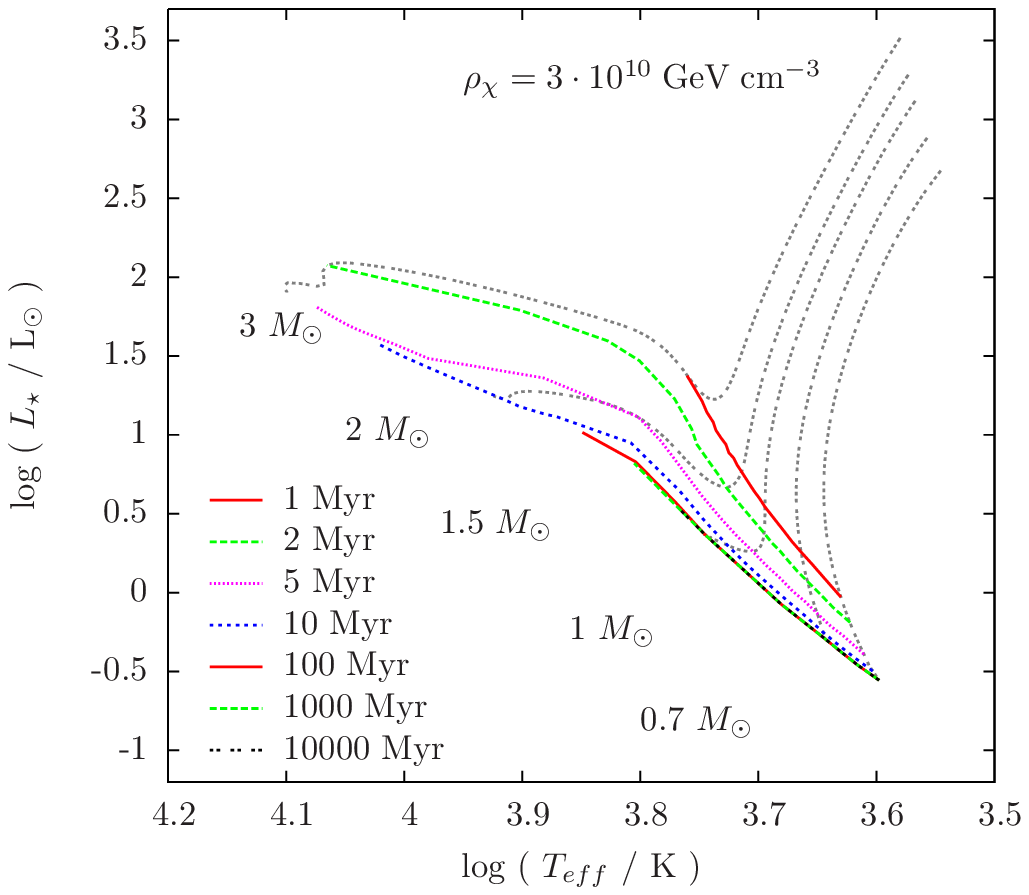}
\vspace{0.1cm} \\
\caption{Tracks on the HR diagram of the collapse of stars of different masses until the ZAMS (grey lines), together with the isochrones of different ages, for DM halo densities $\rho_{\chi} = 0$ and $\rho_{\chi} = 3\cdot10^{10}\;$GeV$\;$cm$^{-3}$. The DM particles are as described in Figure \ref{f_3reg}.}
\label{f_isoHR_coll}
\end{figure} 
\begin{figure}[!t]
\centering
\vspace{0.2cm}
	\includegraphics[scale=0.8]{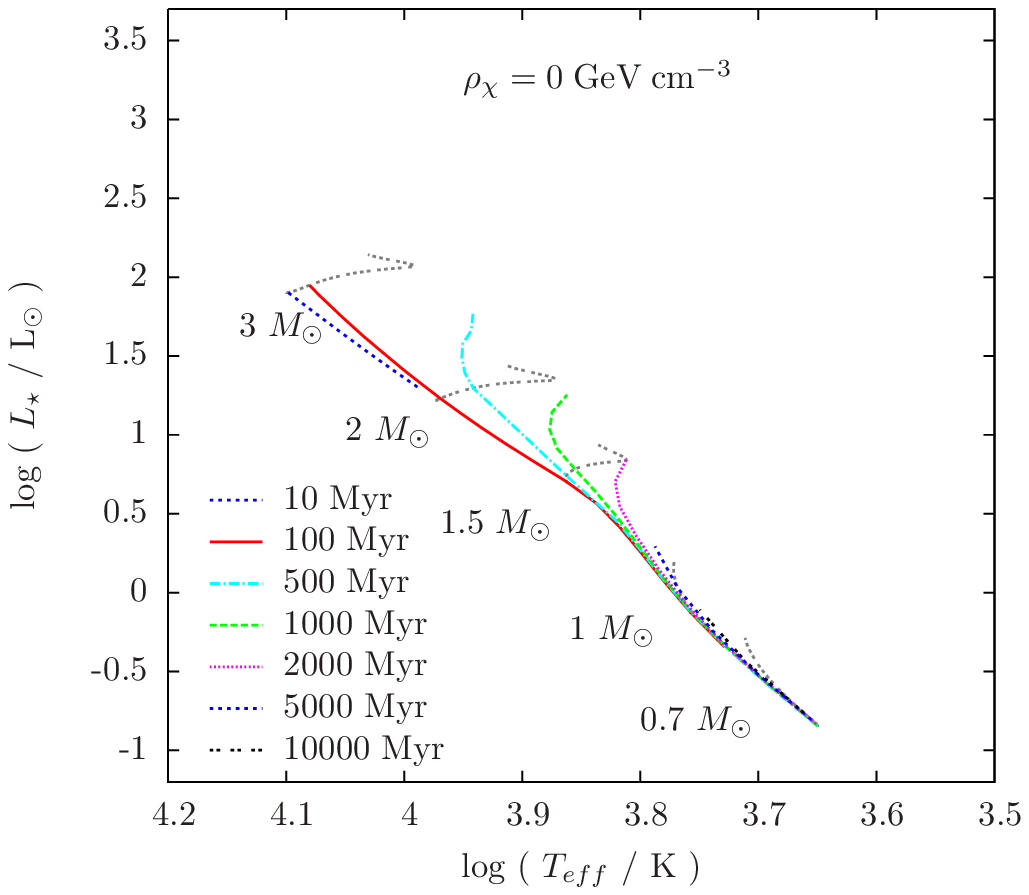}
\vspace{0.7cm} \\
    	\includegraphics[scale=0.8]{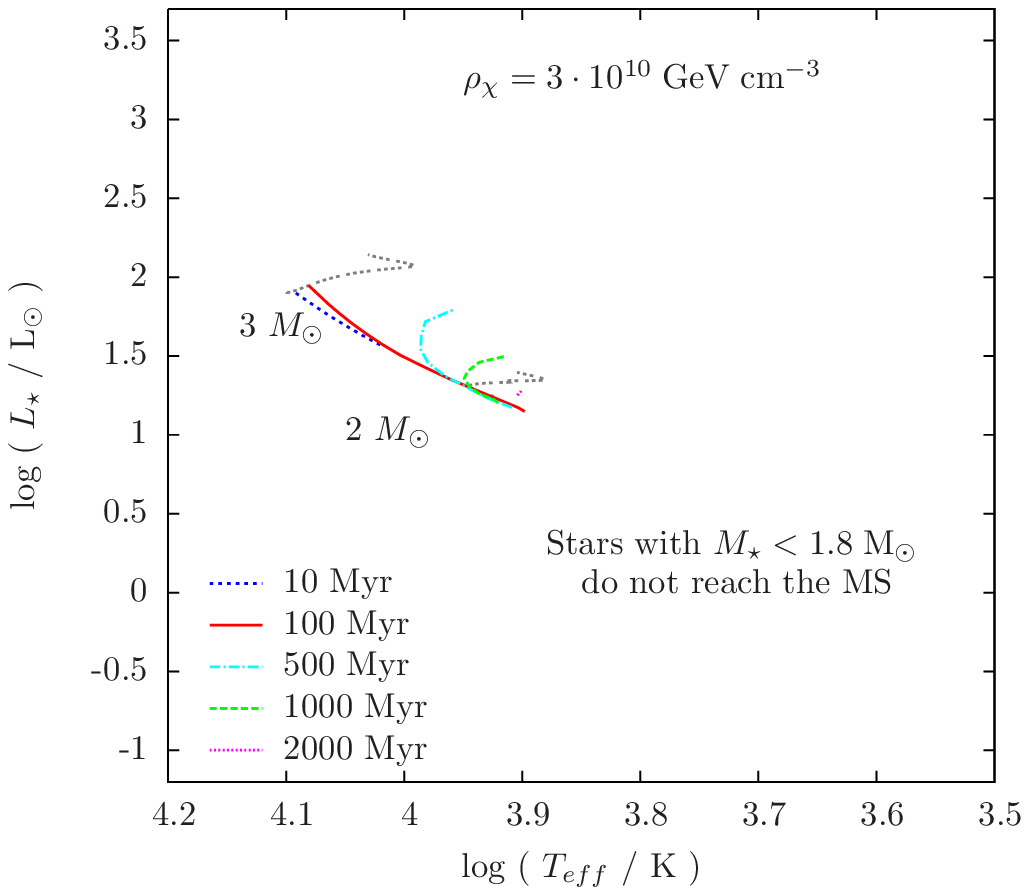}
\vspace{0.1cm} \\
\caption{Tracks on the HR diagram of the evolution through the MS (from $\varepsilon_{grav} < 1\% \varepsilon_{T}$ to $X_{c}<0.001$) of stars of different masses (grey lines), together with the isochrones of different ages, for DM halo densities $\rho_{\chi} = 0$ and $\rho_{\chi} = 3\cdot10^{10}\;$GeV$\;$cm$^{-3}$. The DM particles are as described in Figure \ref{f_3reg}.}
\label{f_isoHR_MS}
\end{figure}
Our approach on the evolution of low-mass stars within DM halos consists in considering the influence of DM capture since the collapse of the star. \citet{art-Ioccoetal2008} did the same to study the evolution of first stars; they also found that the collapse of these stars, the so-called {\it dark stars}, may be stopped at an early stage if there is enough DM on their interior. On the other hand, another approach considered in the literature is to evolve, from the ZAMS, stars that were already formed without DM. This scenario was first analytically estimated by \citet{art-SalatiSilk1989} for main sequence stars and recently numerically simulated by \citet{art-Scottetal2009} for low-mass stars. To compare the two different scenarios, we evolved stars using both approaches and found that they lead to equivalent final equilibrium states, even though the tracks followed by these stars are completely different. When evolved from the ZAMS, stars go back through the pre-main sequence phase, where they reach the same equilibrium states (fuelled only by the energy from DM annihilation) than those obtained when the collapse is \textit{frozen} (see Figure \ref{f_twosides}).

\subsection{Isochrones for low-mass stars evolving in DM halos}

As a synthesis of the new stellar evolution scenarios presented here, we show in Figures \ref{f_isoHR_coll} and \ref{f_isoHR_MS} the paths on the HR diagram followed by stars from 0.7 to 3$\;$M$_{\odot}$ that form and evolve in halos of different DM densities $\rho_{\chi}$, as well as the isochrones of different stages during their evolution. In Figure \ref{f_isoHR_coll} we plotted the collapse until the ZAMS of stars that form in halos with densities $\rho_{\chi} = 0$ and $\rho_{\chi} = 3\cdot10^{10}\;$GeV$\;$cm$^{-3}$. Note that, in the latter case, stars with masses $< 1.8\;$M$ _{\odot}$ completely stop their collapse before reaching the ZAMS (as can be seen from the 10 Gyr isochrone). More massive stars are less affected by DM; their classical evolutionary path is only slightly delayed in a halo with the same DM density. This can also be seen by looking at the 1000 Myr isochrone in Figure \ref{f_isoHR_MS}. In this Figure we plotted the paths of these stars through the MS, that is, from $\varepsilon_{grav}<1\%\;\varepsilon_{T}$ to $X_{c}< 10^{-3}$.
%
\section{Stellar diagnostic on the nature of dark matter particles}
\label{sec-Stellar}
\begin{figure}[!t]
\centering
\includegraphics[scale=0.8]{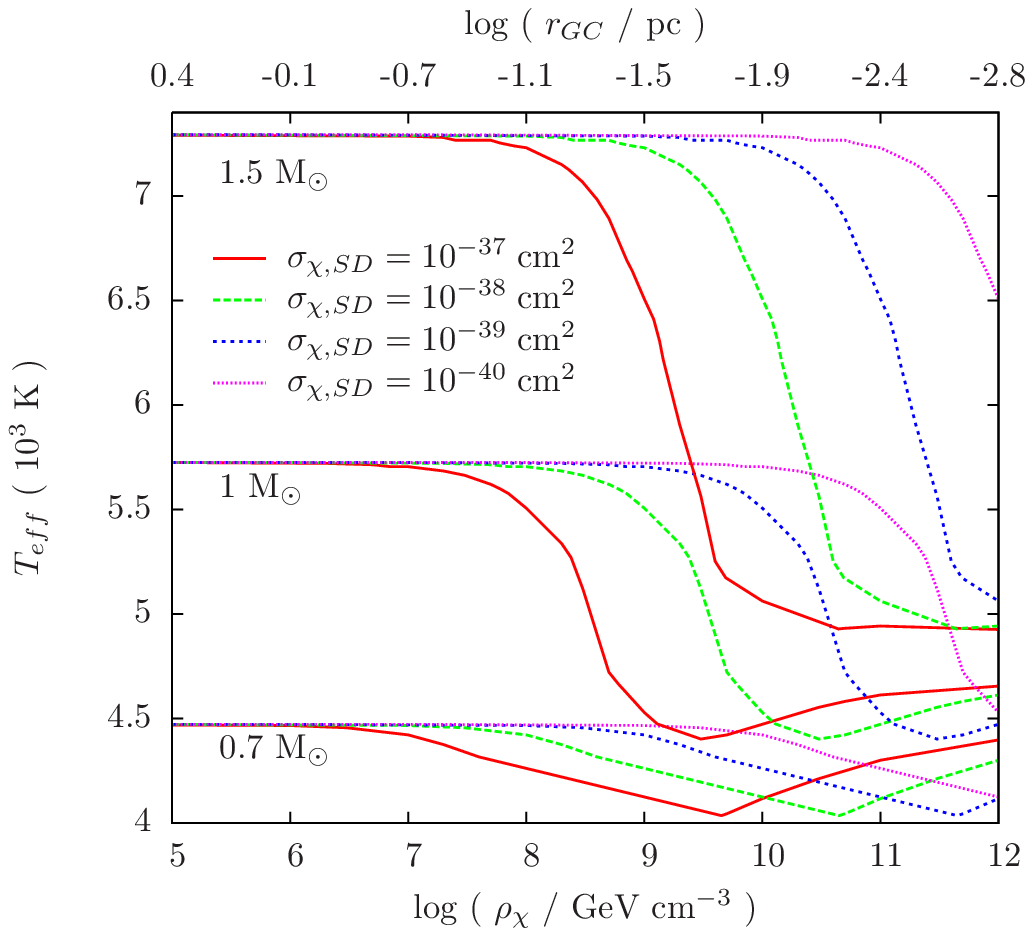}
\vspace{0.3cm} \\
\includegraphics[scale=0.8]{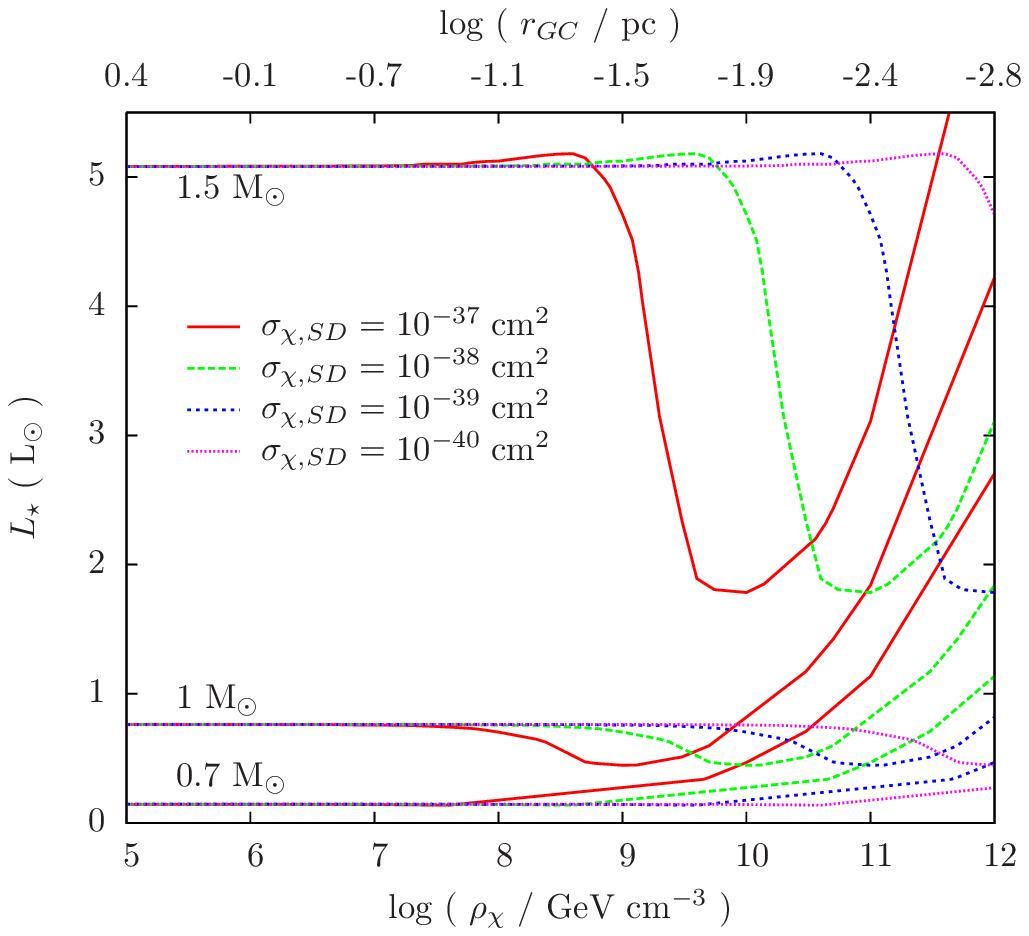}
\caption{Effective temperature (a) and luminosity (b) of stars of 0.7, 1 and 1.5 M$_{\odot}$ as a function of the dark matter halo density, considering different WIMP-proton spin-dependent scattering cross sections. At the top horizontal axe of each figure we show at which distance to the galactic center are these DM densities expected to be found, following the profile of \citet{art-BertonteMerritt2005}. The DM particles are as described in Figure \ref{f_3reg}.}
\label{f_Teff-Lum}
\end{figure}
We have computed luminosity and effective temperature of stars in all scenarios for different values of spin-dependent cross section $\sigma_{\chi,SD}$ and DM density $\rho_\chi$, which we varied from $10^5\;$GeV$\;$cm$^{-3}$ up to $10^{12}\;$GeV$\;$cm$^{-3}$. The results are displayed in Figure~\ref{f_Teff-Lum}, which also shows the possible location of these stars towards the centre of our Galaxy, where the highest densities are expected \citep{art-NavarroFrenkWhite1996, art-Salvador-Sole2007}. In that region, the DM density $\rho_{\chi}$ may be enhanced due to the presence of a supermassive black hole \citep{art-GondoloSilk1999}. We computed the DM distribution around the center of the Milky Way using the adiabatic contracted profile of \citet{art-BertonteMerritt2005}. The predicted T$_{eff}$ and L$_{\star}$ in Figure \ref{f_Teff-Lum} may be used as an alternative method to constrain the WIMP-proton SD scattering cross section $\sigma_{\chi, SD}$, to help in the validation or rejection of DM particles' models, as well as to infer indirectly the DM density $\rho_{\chi}$ in the place where the star is observed. It is worth noting that, at the present moment, these results offer a qualitative picture more than an exact approach, due to the uncertainties in our knowledge of the inner region of our galaxy. In addition to the density profile $\rho_{\chi}(r_{GC})$, both the velocities of the star and of the DM particles also play an important role when studying the stars at the galactic center. \citet{art-Scottetal2009} did precise simulations of possible orbits of low-mass stars in that region, and found that only those stars with elliptical orbits are efficient at capturing DM particles. 

In Figure \ref{f_Teff-Lum} are plotted the $T_{eff}$ and $L_{\star}$ at such an age that all the stars are already in energy equilibrium. Stars that evolve on DM halos of low densities (Weak scenario) are in equilibrium in the beggining of the MS. As $\rho_{\chi}$ increases, the curves mimic a slower evolution through a classical evolution track on the HR diagram (see Figures \ref{f_isoHR_coll} and \ref{f_isoHR_MS}). For high $\rho_{\chi}$ (Intermediate and Strong scenarios), stars are in equilibirum, powered only by the energy from DM annihilation. The higher the value of $\rho_{\chi}$, the sooner the star will freeze its position on the HR diagram at a lower effective temperature $T_{eff}$ and a higher luminosity $L_{\star}$ (see Figure \ref{f_statHR}). In the case of 1$\;$M$_{\odot}$ star, the $T_{eff}$ decreases more than 10$^{3}\;$K and the $L_{\star}$ raises up to three times higher than in the MS. The rapid drop in temperature is related to the fact that the star becomes fully convective. Our results are more conservative than the similar ones found by \citet{art-Fairbairnetal2008}, although they predicted the same behaviour of $T_{eff}$ at lower DM densities. Probably, our underestimation could be overcome by increasing the resolution of our code in the very center of the star, where the energy from DM annihilation is produced.

\section{Conclusions}
\label{sec-conclusion}

One of the consequences of the formation of structures in the Universe is the creation of localised regions with high concentrations of DM. The formation of stars in such peculiar neighbourhoods should be quite different from the usual picture of the formation of young stars by gravitational collapse. 
A striking observational case of such high density DM regions is the young stellar formation regions near the super-massive black holes, located in the centre of galaxies, such as our own Milky Way \citep{art-Genzeletal2003, art-Krabbeetal1995}. In the attempt to grasp the formation of young stars in such unexpected neighbourhoods, in this work we have shown some numerical simulations of stars with masses from 0.7 to 3$\;$M$_{\odot}$ evolving within halos with high density of annihilating DM particles. 

\begin{figure}[!t]
\centering
\begin{tabular}{c c}
\put(55,-2){0.7$\;$M$_{\odot}$}
\put(175,-2){1$\;$M$_{\odot}$}
\hspace{-0.7cm}
\includegraphics[scale=0.5,angle=270]{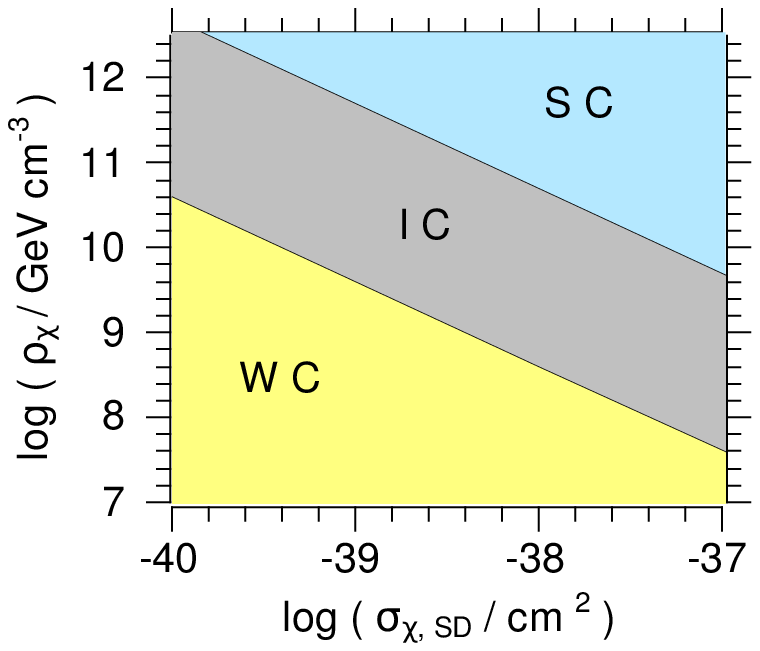} &
\hspace{-1.6cm}
\includegraphics[scale=0.5,angle=270]{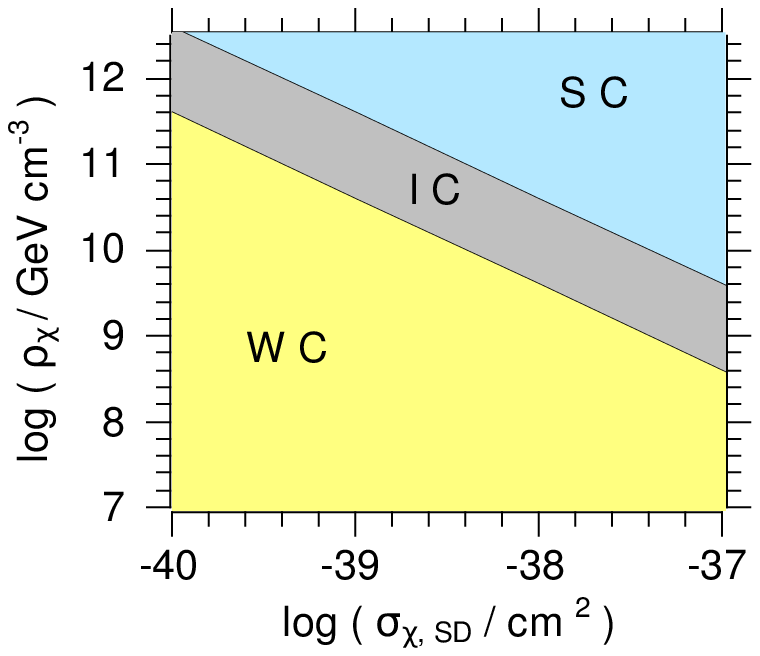} \\
\put(55,-2){1.5$\;$M$_{\odot}$}
\put(175,-2){3$\;$M$_{\odot}$}
\hspace{-0.7cm}
\includegraphics[scale=0.5,angle=270]{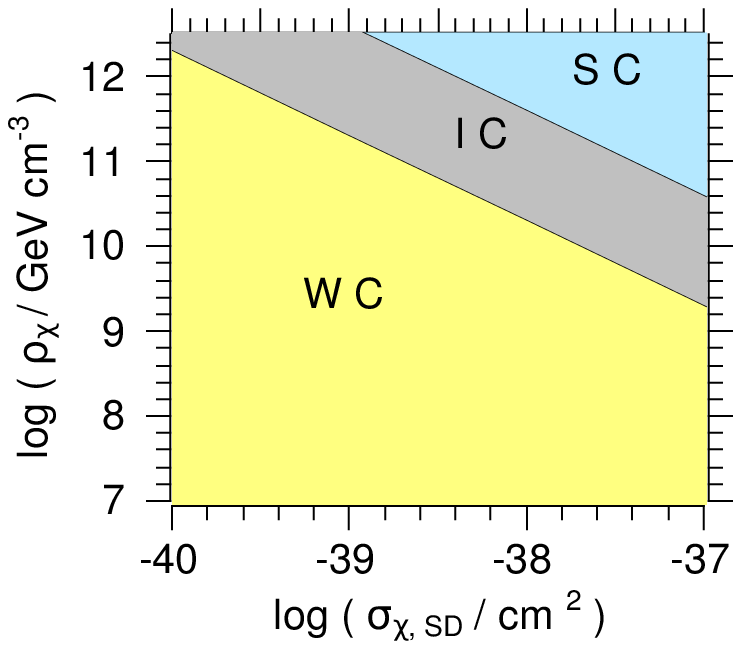} &
\hspace{-1.6cm}
\includegraphics[scale=0.5,angle=270]{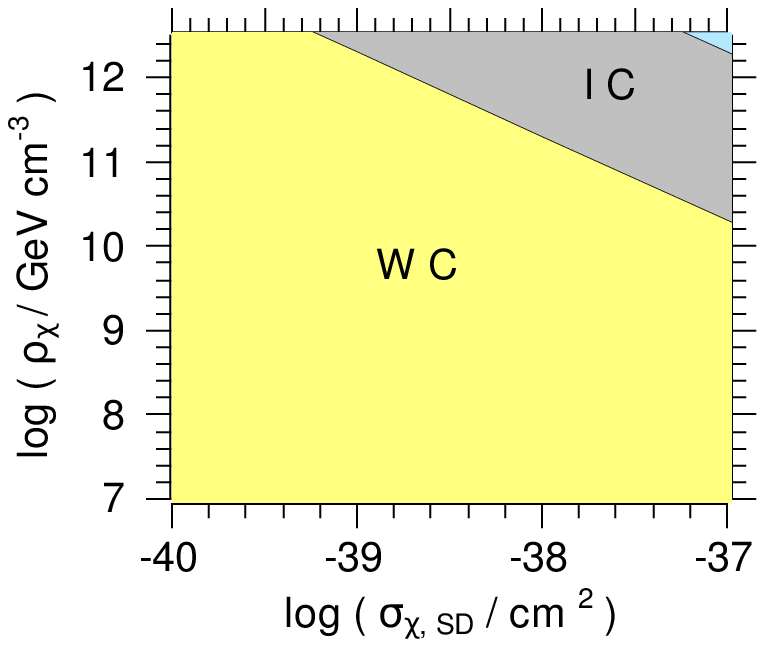} \\
\end{tabular}
\caption{Classification of the different cases of stellar evolution within dark matter halos for stars of 0.7, 1, 1.5 and 3$\;$M$_{\odot}$. Labels \textit{W C, I C} and \textit{S C} indicate, respectively, that these stars are expected to evolve in the Weak, Intermediate or Strong scenarios, considering halos with different DM densities and WIMP-proton spin-dependent scattering cross sections. The DM particles are as described in Figure \ref{f_3reg}.}
\label{f_Class}
\end{figure}

We have found that the evolution of a young star can be affected slightly, moderately  or strongly depending on the DM density of the host halo. Conveniently, we chose to classify the formation and evolution of low-mass stars in three major possible evolution scenarios: Weak, Intermediate and Strong, which are directly related to the amount of DM density. The evolution of the star also depends on the scattering cross section of the DM particles. Figure~\ref{f_Class} shows in which of these three stellar evolution scenarios should stars of 0.7, 1, 1.5 and 3$\;$M$_{\odot}$ evolve, for different values of the SD scattering cross section and the DM density in the halo.

We have shown that low-mass stars in the Strong and Intermediate cases evolve quite differently from the classical path of stars on the HR diagram. During their pre-main sequence phase, these stars will stop their collapse before reaching enough central temperature to start thermonuclear reactions, and will remain indefinitely in the same position on the HR, fuelled only by the energy from WIMP annihilation. In particular, stars immersed in high density DM halos have their effective temperature and luminosity strongly affected due to the change in the energy transport in their interior (Cf. Figure \ref{f_Teff-Lum}).

The new data obtained by means of the near-IR instrumentation allowed the observation of stars in the inner parsec of our own galaxy \citep{art-LuGhezetal2009, art-Ghezetal2005, art-EisenhauerGenzeletal2005}. The observations have revealed a population of apparently young stars in this region, whose current conditions seem to be unsuitable for star formation. This stellar population is usually considered to be a population of old stars that have followed quite an atypical evolution path. If some of these stars were found to be low-mass stars, they would become candidates for this new population of stars evolving in DM halos, as initially suggested by \citet{art-MoskalenkoWai2006}.

If found, such stars would be interesting probes of DM particles near super-massive black holes. Their luminosity, or rather their excess of luminosity, attributed to WIMPs' burning, can be used to derive the WIMPs' matter density at their location. On the other hand, the lack of such unusual stars may provide constraints on WIMPs' density, WIMP-nuclei scattering and pair annihilation cross section. 

Finally, it is worth mentioning that more detailed studies should be done aimed at testing the validity of our model against new stellar observations. This paper should set the foundations for further works.
\acknowledgments
We acknowledge the anonymous referee for his useful comments. This work was supported by a grant from "Funda\c c\~ao para a Ci\^encia
e Tecnologia" (reference numbers POCTI/FNU/50210/2003 and SFRH/BD/44321/2008).

 
\bibliography{DM}

\end{document}